\documentclass[a4paper, 10pt, conference]{IEEEtran}      

\IEEEoverridecommandlockouts                             
\usepackage[utf8]{inputenc}
\usepackage[T1]{fontenc}
\usepackage{graphicx}
\usepackage{hyperref}
\usepackage{caption}
\usepackage{}

\title{\LARGE \bf
Game-Based Discovery: Harnessing Mini-Games within Primary Games for Scientific Data Collection and Problem Solving
}

\author{\IEEEauthorblockN{Abhishek Phadke}
\IEEEauthorblockA{Department of Computer Science\\
Texas A\&M University Corpus Christi\\
Corpus Christi, TX, USA\\
aphadke@islander.tamucc.edu}
\and
\IEEEauthorblockN{Mamta Yadav}
\IEEEauthorblockA{Department of Computer Science\\
Texas A\&M University Corpus Christi\\
Corpus Christi, TX, USA\\
mamta.yadav@tamucc.edu}
\and
\IEEEauthorblockN{Stanislav Ustymenko}
\IEEEauthorblockA{School of CARDS\\
Saint Leo University\\
Saint Leo, FL, USA\\
stanislav.ustymenk@saintleo.edu}
}

\begin{document}

\maketitle
\thispagestyle{empty}
\pagestyle{empty}

\begin{abstract}

In the popular video game Batman: Arkham Knight, produced by Rocksteady Studios and released in 2015, the primary protagonist of the game is Batman, a vigilante dressed as a bat, fighting crime from the shadows in the fictitious city of Gotham. The game involves a real-world player who takes up the role of Batman to solve a peculiar side mission wherein they have to reconstruct the clean DNA sequence of a human and separate it from mutant DNA to manufacture an antidote to cure the villain. Although this is undoubtedly a fascinating part of the game, one that was absent in previous Batman games, it showcases an interesting notion of using mini-games embedded within primary games to achieve a particular real-world research objective. Although the DNA data used in this case was not real, there are multiple such instances in video games where mini-games have been used for an underlying motive besides entertainment. Based on popular case studies incorporating a similar method, this study characterizes the methodology of designing mini-games within primary games for research purposes into a descriptive framework, highlighting the process's advantages and limitations. It is concluded that these mini-games not only facilitate a deeper understanding of complex scientific concepts but also accelerate data processing and analysis by leveraging crowd-sourced human intuition and pattern recognition capabilities. This paper argues for strategically incorporating miniaturized, gamified elements into established video games that are mainly intended for recreational purposes.

\end{abstract}

\section{INTRODUCTION}

In the past five years, players of the video game Borderlands 3, a popular sequel in video game culture, have encountered a peculiar new feature on their "Homeship player base." Within a small section of the in-game environment, a computer terminal was discovered, initiating a new simulation environment in which players' in-game avatars interacted. Ingeniously integrated into the main storyline as a side mission, this mini-game prompted players to dedicate some of their game-playing time to solve a specific challenge presented within the mini-game. Involving the manipulation of Tetris-inspired colored blocks, the mini-game collected players' responses, with these colored blocks representing data from the human gut. The outcomes of players' interactions contributed to an open-source project known as the American Gut Project \cite{grieneisen2018crowdsourcing}. 

This game serves as just one instance of a recent trend: the gamification of research and data collection methodologies. Such games effectively convert complex scientific challenges into engaging tasks that harness the collective problem-solving skills of a diverse group of participants. For example, in biochemistry, mini-games like Foldit enable players to contribute to protein folding studies \cite{cooper2010predicting}. 

Although video games have long been accepted and widely established as a form of entertainment across the globe, appealing to diverse audiences, researchers have also recognized their educational benefits. This recognition has led to studies exploring and developing educational games as effective tools for learning \cite{arias2014using} and knowledge dissemination \cite{stott2013analysis}. However, the scope of "Educational games" extends beyond this. Figure 1 provides a brief categorization of the various uses and forms of games for educational and research purposes.

\begin{figure}[h]
  \centering
  \includegraphics[width=\linewidth]{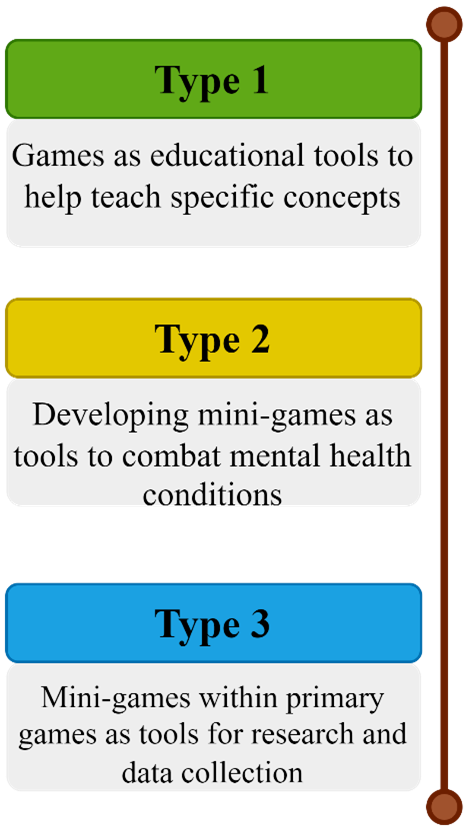}
  \caption{Types of educational games}
\end{figure}

Type 1 games are designed mainly to explain concepts to the audience. For example, a game may be designed for kids where they learn different shapes, colors, or the alphabet. A game may be designed for middle school students, where they can see an interactive display of the human body and learn to place the body organs in the correct positions. Games have been created to explain higher-level concepts to the targeted student audience \cite{smaldone2017teaching}. 

Type 2 games involve games that help with memory, pattern recognition, association, and other essential features of the human mind. These games are usually designed to combat medical issues such as dementia \cite{sea_hero_quest}, vision loss \cite{waddington2015participatory}\cite{gao2018effectiveness}, hearing\cite{melo2023use}, or perception \cite{boot2011do}. 

Type 3 games are not independent games. These are usually mini-games embedded within primary video games that are created for the commercial market. These games offer the player a unique opportunity to participate in crowd-sourced data collection and scientific problem endeavors that the game developers usually have created with an associated research organization. While participation may be purely voluntary, it allows the audience to be a part of a larger scientific community working towards a goal. 

Methodologically, this paper examines the design, execution, and outcomes of these Type 3 mini-games, assessing their effectiveness in generating valid scientific data and insights. Additionally, it addresses the technical and ethical considerations involved in integrating gamification into scientific research. It discusses the balance between data integrity and participatory inclusiveness. Throughout the combined history of scientific research and video game development, there have been multiple instances of a successful collaboration between the two. However, Type 3 games are relatively new, and existing established taxonomies of educational games \cite{o2001taxonomy} may need updating to incorporate them accurately.

\begin{figure*}[h]
  \centering
  \includegraphics[width=\linewidth]{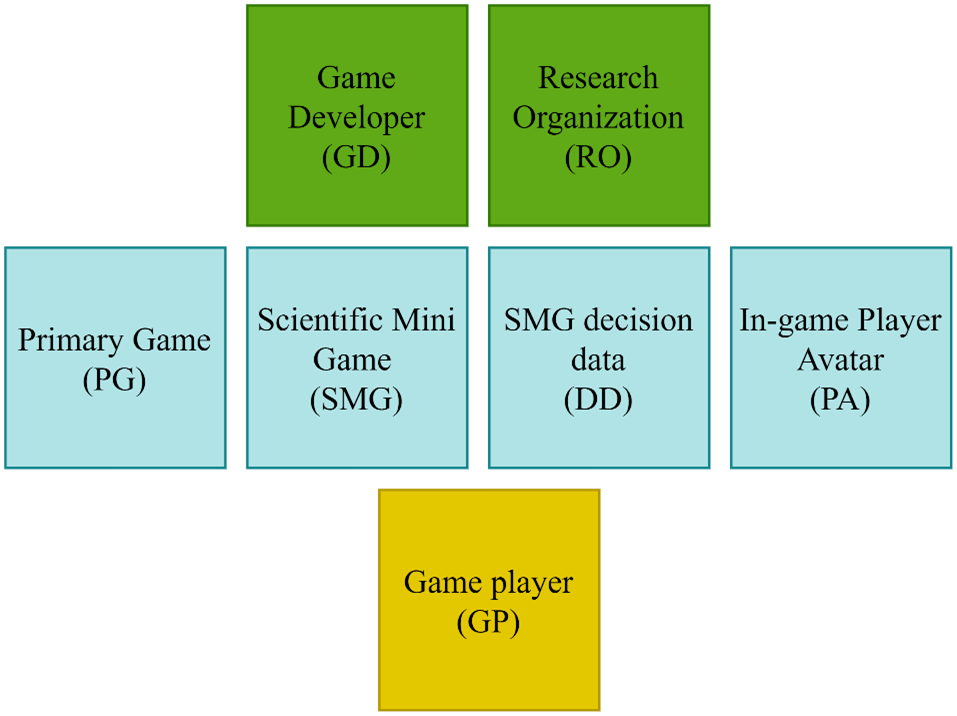}
  \caption{Components described in this study.}
\end{figure*}

One prominent example is Foldit \cite{cooper2010predicting,foldit}, an online puzzle platform initially released in 2008 for protein folding. The game was developed by multiple organizations in collaboration, primarily led by the University of Washington. 
The game aims to challenge gamers by testing their ability to fold protein structures with precision. It provides tools for players to interact with and manipulate the protein structures, aiming to produce optimal results.

Researchers analyzed the highest-scoring results to understand if the solution is a native state that can be applied to the real-world protein. This information helped researchers understand if the solutions can be applied to issues such as diseases. In 2008, the platform had 240,000 registered players \cite{marshall2012online} which shows participation far greater than any similar research survey or crowd-sourced scientific data collection request. 

The scientific publication produced on the process and the game has concluded that a large percentage of the registered players provided relevant results or "outperformed algorithmically computed solutions" \cite{cooper2010predicting}\cite{foldit}. Subsequent published studies have continued to highlight various achievements of the game, remaining pertinent over the past five years \cite{khatib2019building} \cite{koepnick2019de}. 

Sea Hero Quest \cite{sea_hero_quest} was a popular mobile game designed by the British game developer Glitchers in collaboration with associated Alzheimer's disease research centers. The plot of the game involves a sea journey taken by the protagonist, whose role is assumed by the player to navigate, shoot flares, and chase enemies in the game. Data collected from the game helped researchers understand the process of three-dimensional navigation, which is one of the first abilities a person with dementia loses. 

Among the multiple such examples is “Stall Catchers”, which used citizen researchers' help to review footage of blood flow in mice brains to detect Alzheimer's symptoms \cite{stallcatchers}. As illustrated in these case study examples, the primary aim behind game development remained consistent: to engage a broad audience and gather a substantial dataset for researchers to analyze and understand. Although these games were specifically crafted as tools for collecting data to support scientific research, contemporary approaches share similar objectives.

Figure 2 outlines the primary components of the process outlined in the study and briefly describes how they relate to each other. This study distinguishes between the two game subjects as "Primary games" (PGs) and "Scientific Mini games" (SMGs). Primary games are the usual multi-platform video games produced by game studios and distributed and managed by them under various licensure agreements. These games have a global market, are often created for multiple platforms, and have features such as online multi-person cooperative gameplay, expansion packs, and downloadable content packs to enhance and extend gameplay and support for game performance and improvement through further updates over a few years. 

The SMG is a scientific endeavor planned and created by the game studio and a research organization. A game in this category has a specific goal in mind and involves the "Game Player," who is the real-world entity playing the game, using their "In-game Player Avatar" or PA to interact with the SMG in a way similar to the PG. The result of this interaction produces "Decision data" that is then used by the Research Organization to accomplish their scientific goals.

\section{Descriptive Methodology}
In the sections above, this study introduced the reader to the different classifications of educational games and the major components involved in the process. This is followed by a descriptive methodology that can be used to integrate SMGs in PG.

\begin{figure*}[h]
  \centering
  \includegraphics[width=0.8\textwidth]{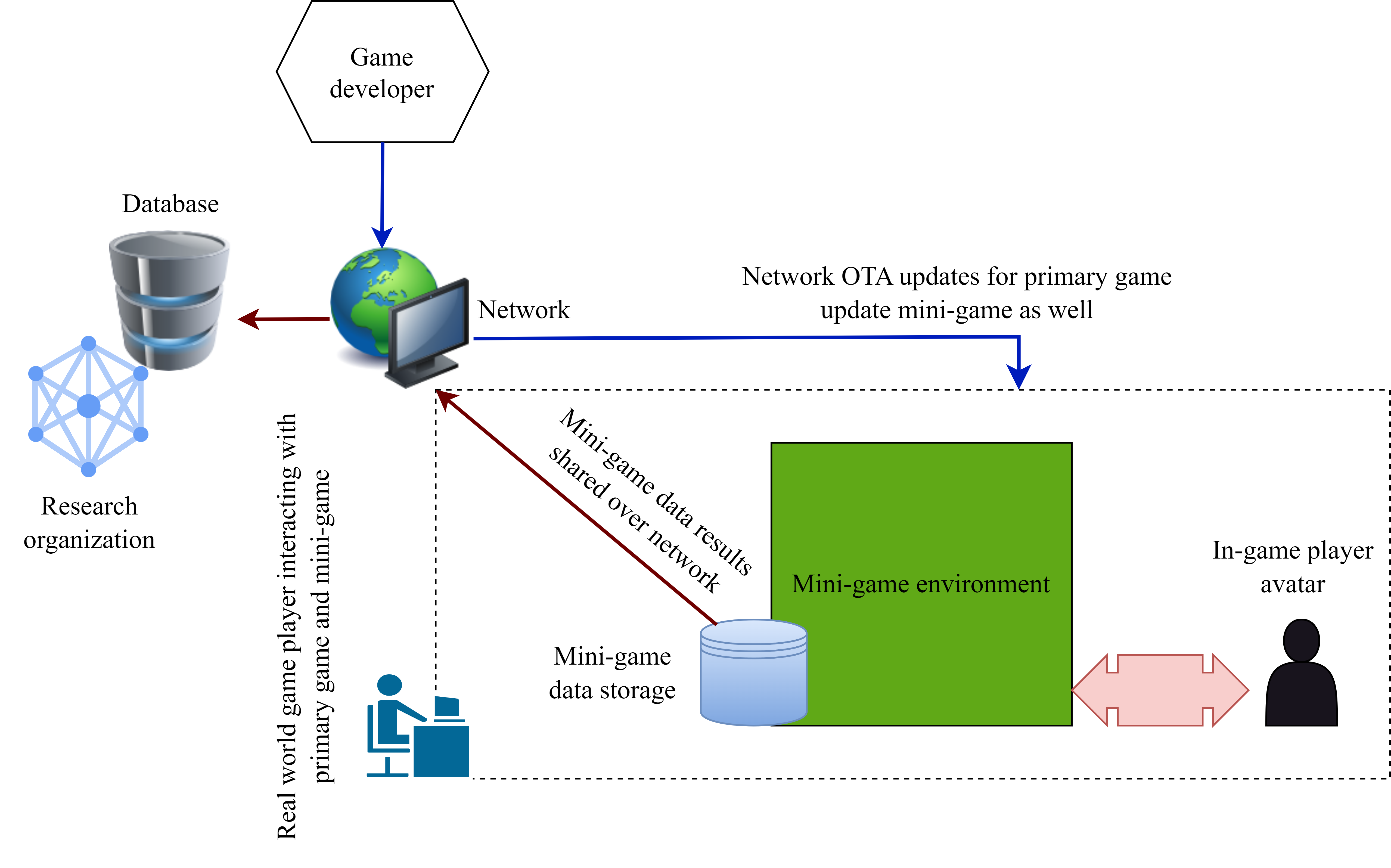}
  \caption{Methodology for  SMGs integration and workflow with respect to Primary games.}
\end{figure*}

Figure 3 shows the descriptive methodology for integrating SMGs into Primary games. The process follows: The SMG is present within the primary game environment and included as a side mission in the primary game's storyline. Typically, the real-world player establishes an in-game avatar to play the primary game. This same avatar is used to interact with the mini-game as well. The mini-game may involve asking the game player to use their player avatar to perform actions such as solving a puzzle, answering specific questions asked by NPC (Non-Playable Characters), or completing other kinds of tasks that are physical in nature for the in-game player avatar. For the real-world game player, this would mean controlling the in-game avatar using the game platform's I/O device. Depending on the data being collected or analyzed, the tasks may differ from the primary game's overall gameplay, style, and intent. The results of this interaction will typically be stored as a separate data structure within the primary game mechanism to protect the data. 

As outlined above, this data is typically labeled as Decision data and includes information about the game players' interaction with the SMG. This decision data is then communicated over the network to the database of the research organization that is collecting the data. The research organization may then analyze the data according to its goals. Any update to the SMG is done through the game developer, who will push an OTA update to the primary game through established channels. In some instances, a focused update to modify the contents of the SMG can also be performed.

\begin{table*}[h]
    \centering
    \begin{tabular}[width=\linewidth]{|p{0.3\linewidth}|p{0.3\linewidth}|p{0.3\linewidth}|}
        \hline
        \textbf{Process} & \textbf{Example} & \textbf{Expected data generated} \\
        \hline
        Show in-game pictures to the PA and ask them to type/choose the correct object. 
        & The PA is shown pictures of a 'Tree,' a 'Car,' and an 'Apple' and asked to tag the 'Apple' object. 
        & Will produce labeled data that can be used to train real-world ML and Computer Vision algorithms \\
        \hline
        Moving physical items that are represented by game models in the PG  
        & The PA is asked to construct a tower using the given pieces of wood in-game 
        & Understand how humans perceive vital knowledge such as architecture, using resources and physics \\
        \hline
        Examine biological sequences such as cell structures and protein lengths 
        & The PA is asked to construct different sequences of cells, protein information, or other biological data 
        & Allows the researchers to understand how an audience can construct or reconstruct particular target structures and get this data for further processing. \\
        \hline
        Construct permutations of variable lengths 
        & The PA is asked to take particular objects in the game and arrange them in patterns or permutations. 
        & This allows the researchers to get a large amount of permutation data for the target object being studied, such as DNA or RNA sequences. \\
        \hline
        Study reaction times for diverse audiences 
        & The PA is asked to input the age of their GP and then made to shoot/follow/catch particular objects in-game at variable speeds and times. 
        & Pseudo-health data, such as player reaction times, avoidance decisions, or perception of stimuli, is checked. \\
        \hline
        Uses the in-game mechanics to study player interactions with the game on specific targets, such as planning or optimization & Tactical and strategy city-builder games examine how players build/manage economies. & Data determines how players understand concepts in politics, finance, and management. \\
        \hline
    \end{tabular}
    \caption{In game process, example, and expected data generated from SMG integration into Primary games.}
    \label{tab:data}
\end{table*}

Table 1 highlights how SMGs can be integrated into PGs, the process they would follow, an example implementation, and the expected data generation.

\section{Challenges and limitations}
This section discusses the various challenges of implementing mini-games as viable tools for data collection. Starting with the challenges to integration and maintenance of SMGs within the primary game environment, several issues will need to be addressed on an ongoing basis for the research organization to maintain effective relationships with the game developer and the player.

One of the significant goals of integration outlines that the mini-game participation should be optional and not a part of the main storyline. While this is done to prevent degrading the player's gameplay experience, the success of the data collection or procuring player information from the mini-game interaction depends mainly on the player's choice to interact with the SMG. If the player chooses to skip the SMG entirely and continue with the primary gameplay storyline only, no information can be collected. 

Even if the decision to integrate the SMG has been made at the initial stage of game design and development, the creation of the SMG itself, along with its interactive elements placed within the game, and other aspects of game design such as crafting appropriate assets, sound files, and rewards, require time and resources. These efforts may ultimately incur costs for the developer. Additionally, since the SMG operates as a piece of code executed within the primary game environment, it must undergo the same rigorous testing and verification procedures before deployment. A glitch in the SMG code could potentially disrupt or even cause the primary game to malfunction, which would not only impact the game development process but also damage the reputation of the game developer and its product.

Most offline games designed these days do not require a persistent, active Internet connection to be maintained by the game console to make the game playable. In such scenarios, explicit permission must be obtained from the user for the console and the game to be allowed to connect to the internet and share the decision data obtained by the mini-game.

As data collection advances and the needs of the managing RO change, a parallel change in the SMG may be required. While this can be achieved through an update, in some cases, it might involve pushing an update through the primary game, which may be tedious. Also, most updates to the software running on proprietary gaming consoles must be verified independently by the hardware providers, which may add an extra facet of time before the new version is released. Since the entire process is tedious and resource-consuming, it is only logical to implement it in games and game-development studios with a huge fan following and commercial market. Smaller indie developers may not have the time or capability to manage the integration of SMGs in initial builds. However, this issue can be easily solved over slow, long-term releases in later years.

Validating the results obtained through crowd-sourced data labeling, annotation, or even decision-based results may require additional vetting from the research organization. This is necessary to examine outliers, prevent errors, and detect additional patterns in the data obtained. A similar challenge would be disseminating the results produced using these methods, as the data is collected over such a wide audience track over extended periods that it might be impossible to establish milestones in data collection and processing. As such, the academic and scientific community may bear additional skepticism about the veracity of the collected information without sufficient proof. Fortunately, any additional scrutiny can be passed by creating a documented and streamlined data collection and analysis process.

\section{Advantages of incorporation of SMGs in PGs}
\begin{figure*}[h]
  \centering
  \includegraphics[width=0.9\linewidth]{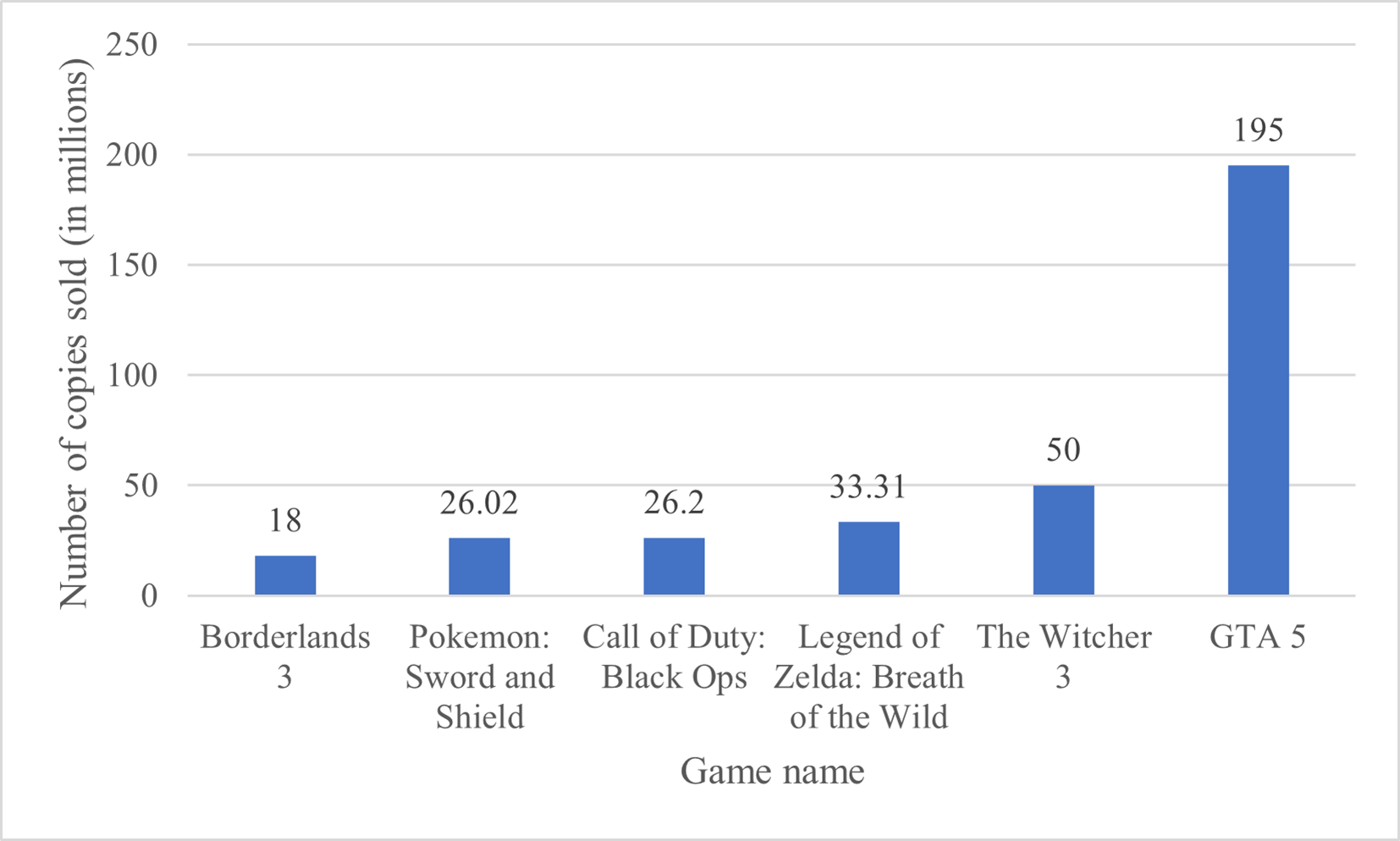}
  \caption{Popular games categorized by the number of copies sold worldwide in millions by 2023.}
  \label{fig:games_sold}
\end{figure*}

The following subsection discusses the advantages of including SMG in primary games as an effective data-gathering and problem-solving method. There is a vast and ever-growing market for games, from popular established GDs sequels to established games, remakes, and new games in all genres. This market is many times bigger in terms of reciprocity than current research surveys get. Figure \ref{fig:games_sold} lists popular games and the millions of copies sold worldwide.

Indeed, suppose one were to imagine capturing just a tiny percentage of the audience from this market willing to interact with the SMG as they play the primary game. In that case, the possibility of capturing data and problem-solving using distributed means is endless and spans a global range. Most puzzle-solving, strategy, and RPG (Role Playing Games) already put the players in a similar mindset where they solve various challenges in-game to accomplish a particular goal. As such, the player in this state is a perfect audience for such SMGs to target and gain valuable information from. 

It has been realized that most, if not all, survey and crowd-problem-solving endeavors require that the user be compensated for their time or incentivized to participate in some way. While actual monetary compensation may be difficult to accomplish because of obvious income, taxation, and misuse prevention regulations, it is very easily possible to allow the players to be awarded in-game currency, perks, and collectible items for completing milestones dictated by the SMG within the primary game. Sometimes, even with monetary compensation involved, collecting opinions and data from the public often has fewer responses. Adding mini-game approaches in-game might provide a richer data stream. Based on specific case studies, this approach has been done before with a reasonable success rate in Borderlands 3, as described above \cite{amsen2024millions}.

Additionally, incorporating these SMGs will not affect the overall gameplay experience, as completing them is voluntary. Skipping or completing these SMGs will not affect the game's central plot or the player's ability to complete the primary game. A certain group of game players make it their goal to acheieve a 100\% game completion achievement. Skipping the SMG will not affect the percentage completion rubric of the PG. These games merely serve as side quests for the player to complete should they wish to take a break from primary gameplay or simply for them to return to the game once they have completed all the missions in the primary game. It is also unlikely that incorporating these SMGs will increase the overall file size of the game or affect gameplay performance to a noticeable extent. 

Since most games have a relatively global market with fewer restrictions than other forms of data collection, a wider audience can be reached. By default, GDs include language pack downloads with their games that enable them to be played by a broader range of audiences. Integrating SMG would harness the current existing capabilities of primary games and their geographical reach to provide greater diversity to the data and decisions obtained. Often, the primary games are played by enthusiastic followers and fans over decades, enabling the SMGs present within them to have a more prolonged and persistent presence that would reach a wider generation audience over an extended period. This has other advantages, too. If the needs of the research study supported by the SMG change with time, it is a matter of updating the mini-game to accommodate the new requirements for data collection, questionnaires, or target experiments. This change can be pushed as an update to the primary game or simply a focused refresh to update the SMG. It is logical to assume that players are more likely to interact with mini-game surveys and data collection exercises than with current methods of focus groups, phone interviews, and in-person interactions. This is especially relevant if, as a means of compensation, the SMGs provide the player's in-game avatar with in-game currency, perks, or collectible items.

\section{Future work and conclusions}
Although the advantages of including SMGs in PGs are significant, challenges, as discussed above, still need to be addressed. Research organizations that seek to use such games and integrate them into their data collection or decision-making processes must work towards maintaining a fine balance between helping keep GDs on their development timelines without causing them damage in terms of resources or time spent, keeping the GPs playing experience the same and collecting data in a manner that is acceptable to the scientific community. Additional possibilities include expanding to other platforms through which Video games can be played. With the advent of virtual reality headsets, game developers have started porting many existing games and creating new educational games to cater to the VR market \cite{lampropoulos2024virtual}. SMGs implanted into these games may provide additional benefits as well. In conclusion, the integration of mini-games as agents for scientific problem-solving and data collection, wrapped up as side missions in larger, established primary games, represents a promising frontier in research, offering new pathways for discovery and innovation.

\addtolength{\textheight}{-12cm}

\bibliographystyle{IEEEtran}
\bibliography{main}

\end{document}